\documentclass{article}
\usepackage[cp1250]{inputenc}
\usepackage[T1]{fontenc}
\usepackage[english]{babel}
\usepackage{amsmath}
\usepackage{amssymb}
\usepackage{hyperref}
\usepackage{url}
\usepackage{calc}
\numberwithin{equation}{section}

\begin{document}

\title{On the Empirical Importance of the Conditional Skewness Assumption in Modelling the Relationship Between Risk and Return
\thanks{Presented at the $3$-rd Symposium on Socio- and Econophysics, FENS2007, Wrocław $22-24$ November $2007$. To be published in Acta Physica Polonica A}
}
\author{Mateusz Pipień\\
Department of Econometrics, Cracow University of Economics\\ 
Rakowicka 27, 31-510 Kraków \\
e-mail: \texttt{eepipien@cyf-kr.edu.pl}}
\date{ }

\maketitle

\begin{abstract}
We present the results of an application of Bayesian inference in testing the relation between risk and return on the financial instruments. On the basis of the Intertemporal Capital Asset Pricing Model, proposed in \cite{Mert1973} we built a general sampling distribution suitable in analysing this relationship. The most important feature of our assumptions is that the skewness of the conditional distribution of returns is used as an alternative source of relation between risk and return. This general specification relates to Skewed GARCH-In-Mean model proposed in \cite{OsiPip2000}.
In order to make conditional distribution of financial returns skewed we considered a constructive approach based on the inverse probability integral transformation presented in details in \cite{Pip2006}. In particular, we apply hidden truncation mechanism, two equivalent approaches of the inverse scale factors, order statistics concept, Beta and Bernstein distribution transformations, and also the method recently proposed in \cite{FerrSteel2006}.
Based on the daily excess returns on the Warsaw Stock Exchange Index we checked the empirical importance of the conditional skewness assumption on the relation between risk and return on the Warsaw Stock Market. We present posterior probabilities of all competing specifications as well as the posterior analysis of the positive sign of the tested relationship.\\


\end{abstract}

\section{Introduction}
The basics of the financial economics is constituted by the relationship between risk and return. Numerous papers have investigated this fundamental trade-off testing linear dependence of excess return on the level of risk, both measured by conditional mean and conditional standard deviation of the value of investor's wealth. According to \cite{Mert1973}, given risk aversion among investors, when investment opportunity set is constant, there is a positive relationship between expected excess return and risk. Hence, it is possible to express the risk in terms of the expected premium generated. 
\\Historically, authors have found mixed empirical evidence concerning the relationship. In some cases a significant positive relationship can be found, in others it is insignificant and also some authors report it as being significantly negative. For instance, using monthly U.S. data \cite{FrSchSta1987} and also \cite{CampHent1992} found a predominantly positive but insignificant relationship. Glosten, Jagannathan and Runkle reported in \cite{GloJagRun1993} a negative and significant relationship on the basis of Asymmetric-GARCH model, instead of commonly used GARCH-in-Mean framework; see \cite{EngLilRob1987}. Scruggs summarises the empirical evidence of considered relationship in \cite{Scr1998}.
\\The main goal of this paper is an application of Bayesian inference in testing the relation between risk and excess return of the financial time series. We revisited Intertemporal Capital Asset Pricing Model (ICAPM) in order to investigate the empirical importance of the skewness assumption of the conditional distribution of excess returns. On the basis of the model, we built a general sampling distribution of the observables suitable in estimating risk premium. The most important feature of our model assumptions is that the possible skewness of conditional distribution of returns is used as an alternative source of relation between risk and return. Thus pure statistical feature is equipped with economic interpretation. Our general specification fully corresponds to suggestion, that systematic skewness is economically important and governs risk premium; see \cite{HarSid2000}. In order to make conditional distribution of financial returns skewed we considered a constructive approach based on the inverse probability integral transformation presented in details in \cite{Pip2006}. Based on the daily excess returns of index of the Warsaw Stock Exchange we checked the total impact of conditional skewness assumption on the relation between return and risk on the Warsaw Stock Market. On the basis of the posterior probabilities and posterior odds ratios, we test formally the explanatory power of competing, conditionally fat tailed and asymmetric GARCH processes.

\section{Creating asymmetric distributions}
The unified representation of the univariate skewed distributions that we apply in this paper is based on the inverse probability integral transformation; for details see \cite{Pip2006}. The family of random variables $IP=\{\varepsilon_s, \varepsilon_s: \Omega\to \mathbb{R}\}$, with representative density  $s(.|\theta, \eta_p)$ is called the skewed version of the symmetric family I (of random variables with unimodal symmetric density $f(.|\theta)$ and distribution function $F$, such that the only one modal value is localised at $x=0$) if $s$ is given by the form:
\begin{equation}\label{density:1}
s(x |\theta, \eta_p)=f(x|\theta)\cdot p\left(F(x|\theta )
|\eta_p\right),\quad x\in\mathbb{R}.
\end{equation}
The asymmetric distribution $s(.|\theta, \eta_p)$, where $\theta$ is inherited from the density $f$ and $\eta_p$ groups the skewness parameters, is obtained by applying the density $p(.|\eta_p)$ as a weighting function. Within the general form (\ref{density:1}) several classes of distributions have been imposed on some specific families of symmetric random variables. A review of skewing mechanisms was presented in \cite{Pip2006}. Here we apply (\ref{density:1}) in order to build the set of unnested specifications, which compete in explaining the possible relationship between risk and return.

\section{Basic model framework and competing skewed conditional distributions}
Let denote by $x_j$ the value of a stock or a market index at time $j$. The excess return on $x_j$, denoted by $y_j$, is defined as the difference between the logarithmic daily return on $x_j$ in percentage points ($r_j =100\ln(x_j/x_{j-1})$) and the risk free short term interest rate (denoted by $r_j^*$), namely $y_j=r_j-r_j^*$. The voluminous literature focused on examination the relationship between risk and return bases on the Intertemporal CAPM, proposed by Merton in \cite{Mert1973}. According to assumptions of Merton's theory there exists a set of distributions $P(.|\psi_{j-1})$, conditional with respect to the information set at time $j$ (denoted by $\psi_{j-1}$) such, that:
\begin{equation}\label{density:2}
E(y_j|\psi_{j-1})=\alpha^{*}D(y_j|\psi_{j-1}),
\end{equation}
where symbols $E$ and $D$ denote expectaction and standard deviation respectively. The coefficient $\alpha^{*}>0$ in (\ref{density:2}) measures the relative risk aversion of the representative agent. Under assumptions of the informational efficiency of the market, the information set at time $j$ can be reduced to the history of the process of the excess return, namely $\psi_{j-1}=(\ldots,y_{j-2},y_{j-1})$. Consequently, an econometric model of the relationship between risk and return should explain the properties of the conditional (with respect to the past of the process $y_j$)  distribution of the excess return $y_j$ at time $j$. It is also of particular interest to find any linkage between expected excess return and the measure od dispersion of the distribution of $y_j$, conditional to $\psi_{j-1}$. Following \cite{EngLilRob1987}, \cite{FrSchSta1987} and \cite{OsiPip2000} we consider for $y_j$ a simple GARCH-In-Mean process, defined as follows:
\begin{equation}\label{model:1}
y_j=[\alpha+E(z_j)]\sqrt{h_j}+u_j,\qquad j=1,2,\ldots,
\end{equation}
where $u_j=[z_j-E(z_j)]\sqrt{h_j}$, and $z_j$ are independently and identically distributed random variables with $E(z_j)<+\infty$. The scale factor $h_j$ is given by the GARCH(1,1) equation; see \cite{Boll1986}:
\[
h_j=\alpha_0+\alpha_1u_{j-1}^2+\beta_1h_{j-1},\qquad j=1,2,\ldots. .
\]
The specific form of the conditional distribution of $y_j$ in (\ref{model:1}) is strictly dependent on the type of the distribution of $z_j$. Initially, in model denoted by $M_0$, we assumed for $z_j$ the Student-$t$ density with unknown degrees of freedom $\nu>1$, zero mode and unit inverse precision:
\[
z_j|M_0 \sim iiSt(0,1,\nu), \qquad \nu>1.
\]
The density of the distribution of $z_j$ in $M_0$ is given as follows:
\[
p(z|M_0)= f_t(z|0,1,\nu) = \frac{\Gamma (0.5(\nu +1))}{\Gamma
(0.5\nu)\sqrt{\pi \nu}}\left[1+\frac{z^2}{\nu}\right]^{-(\nu +1)/2}
\]
Given model $M_0$, $E(z_j)=0$, $u_j=z_j\sqrt{h_j}$, and hence (\ref{model:1}) reduces to simpler form $y_j=\alpha \sqrt{h_j}+u_j$. Let denote by $\theta=(\alpha, \alpha_0, \alpha_1, \beta_1, \nu)$ the vector of all parameters in model $M_0$. Here, the conditional distribution of the error term is the Student-$t$ distribution with degrees of freedom parameter $\nu>1$, zero mode and inverse precision $h_j$. Consequently, the following density represents conditional distribution of the excess return at time $j$:
\[
p(y_j|\psi_{j-1},\theta,M_0)= h_j^{-0.5}f_t(h_j^{-0.5}(y_j-\alpha\sqrt{h_j})|0,1,\nu),j=1,2,\ldots.
\]
Given model $M_0$ the expected excess return (conditional to the whole past $\psi_{j-1}$) is proportional to the square root of the inverse precision $h_j$:
\begin{equation}\label{riskret:1}
E(y_j|\psi_{j-1},\theta,M_0)=\alpha\sqrt{h_j},\quad j=1,2,\ldots.
\end{equation}
The parameter $\alpha\in R$ captures the dependence between expected excess return and the level of risk, both measured by $E(y_j|\psi_{j-1},\theta,M_0)$ and the scale parameter $\sqrt{h_j}$ respectively. 
\\Now we want to construct a set of competing GARCH specicifations $\{M_i,i=1,\ldots,k\}$ by introducing skewness into density of the conditional distribution of excess return, $p(y_j|\psi_{j-1},\theta,M_0)$. The resulting asymmentric distributions are obtained by skewing the distribution of the random variable $z_j$, according to method presented in the previous section. The asymmetric density of $z_j$ is of the general form related to the formula (\ref{density:1}):
\[
p(z|M_i)= f_t(z|0,1,\nu)p[F_t(z)|\eta_i,M_i], z\in R,i=1,2,\ldots,k,
\]
where $p(.|\eta_i,M_i)$ defines the skewing mechanism parameterised by vector $\eta_i$ and $F_t(.)$ is the cdf of the standardised Student-$t$ distribution. This leads to the general form of the conditional distribution of daily excess return $y_j$ in model $M_i$:
\[
p(y_j|\psi_{j-1},\theta,\eta_i,M_i)= h_j^{-0.5}f_t(z^*_j|0,1,\nu)p[F_t(z^*_j)|\eta_i,M_i] ,j=1,\ldots,
\]
where $z^*_j=h_j^{-0.5}(y_j-\mu_j)$ and $\mu_j=[\alpha+E(z_j)]\sqrt{h_j}$.
\\By imposing skewness, the expectation  $E(z_j)$ is no longer equal to zero. Consequently, given $M_i$, the  expectation of the excess return (conditional to $\psi_{j-1}$) is still proportional to $\sqrt{h_j}$, but the coefficient of proportionality changes:
\[
E(y_j|\psi_{j-1},\theta,\eta_i,M_i)=[\alpha+E(z_j)]\sqrt{h_j}.
\]
Hence, the skewness of conditional distribution of $y_j$ is treated in $M_i$ as another source of the tested relationship.
\\As the first specification, denoted by  $M_1$, we consider GARCH model with skewed Student-$t$ distribution obtained by the method proposed in
\cite{FernSteel1998}, or equivalently in \cite{Hansen1994}. The model $M_2$ is the result of skewing conditional distribution according to the hidden truncation idea; see \cite{Azzalini}. In models $M_3$ and $M_4$ we apply Beta skewing mechanism; see cite{Jones2004}. In model $M_5$ we apply Bernstein density based skewing mechanism with $m = 2$ free parameters, while model $M_6$ is built on the basis of the skewing construct defined in \cite{FerrSteel2006}. All competing specifications, together with analytical forms of skewing mechanisms and model specific parameters are presented in Table 1. For some details concerning sampling densities and prior specifications in each model see \cite{Pip2006}.

\section{Empirical results for WSE index}
In this part we present an empirical example of Bayesian comparison of all competing specifications. We also discuss the posterior analysis of the total impact of the conditional skewness assumption on the relationship between risk and return on the Warsaw Stock Exchange (WSE). Our dataset $y$ was constructed on the basis of $t$=2144 observations of daily growth rates, $r_j$, of the WSE index (WIG) from 06.01.98 till 31.07.06. The risk free interest rate, $r_j^*$, used to calculate excess return $y_j$, was approximated by the WIBOR overnight interest rate (WIBORo/n instrument). Our empirical results remained practically unchanged for $r_j^*$ calculated on the basis of the middle and long term WIBOR Polish Zloty interest rates and also in the case $r_j^*=0$.
\\Table 2 presents posterior probabilities $P(M_i|y)$ calculated for each of competing models $M_i$, $i=0,1,\ldots,6$. The initial specification $M_0$, built on the basis of the conditional symmetric Student-$t$ distribution, receives a little data support, as the posterior probability $P(M_0|y)$ is slightly greater than 8\%. All remaining posterior probability mass is attached to specifications which allow for conditional skewness. It is clear, that the modelled dataset of excess returns of WIG index do not support decisively superiority of any of the competing skewing mechanism. The greatest value of $P(M_i|y)$ receives conditionally skewed Student-$t$ GARCH model generated by the Beta distribution transformation with two free parameters. In this case the value of posterior probability is equal about 40\%. The dataset also support conditionally skewed Student-$t$ GARCH model with hidden truncation mechanism ($M_2$) and Beta distribution transformation with one free parameter ($M_3$). Those three models cumulate more than 90\% of the posterior probability mass, making all remained conditionally skewed specifications improbable in the view of the data. Thus, inverse scale factors, the Bernstein density transformation and construct proposed in \cite{FerrSteel2006}, namely models $M_1$, $M_5$ and $M_6$, lead to very doubtful explanatory power. Those specifications are strongly rejected by the data, as the values of posterior probabilities are much smaller than posterior probability of symmetric GARCH model.
\\In Table 2 we also compare the total impact of the conditional skewness effect on the tested relation between risk and return. According to our assumptions, the conditional expectation of the excess return is proportional to the square root of the inverse precision $h_j$. Since we parameterize the market risk by a more general dispersion measure than standard deviation we report the information about the relative risk aversion by the posterior probability of the positive sign of the function $\alpha+E(z_j)$. According to (\ref{model:1}) it enables to test the positive sign of the relative risk aversion coefficient. Initially we checked the strength of the relation in model $M_0$, which does not allow for conditional skewness. Given $M_0$ $E(z_j)=0$ and the whole information about relative risk aversion is reflected in parameter $\alpha$; see (\ref{riskret:1}). Just like many researchers we obtained positive, but rather weak, relation between expected excess return and risk, given model with symmetric conditional distribution. The posterior probability $P(\alpha >0|M_0,y)$, equal about 0.92, leaves considerable level of uncertainty about true strength of the tested relation. Consequently, model $M_0$ does not confirm our hypothesis strongly. Imposing unreasonable (in the view of the data) skewness into conditional distribution of excess returns also may not strengthen our inference. In case of models with weak data support ($M_1$ and $M_6$) the assumption of asymmetry of the density $p(y_j|\psi_{j-1},\theta,\eta_i,M_i)$ does not improve posterior inference about the sign of $\alpha+E(z_j)$. In case of $M_1$ and $M_6$ posterior probability of positive relationship is very close to the value generated within $M_0$. Only in case of the skewing mechanisms with the greatest data support, namely Beta transformation with two parameters and hidden truncation, the WIG excess returns yield decisive support of the positive sign of the relative risk aversion coefficient. In case of model $M_3$, the posterior probability of positive sign of $\alpha+E(z_j)$ is greater than 0.99, leaving no doubt about the significance of the relationship between risk and return postulated by Merton in \cite{Mert1973}. Hence, it was possible to confirm positive sing of $\alpha+E(z_j)$ strongly only by imposing specific skewing mechanism into conditional distribution of excess returns. Beta distribution transformation with two free parameters was able to detect additional source of information about risk premium in the WIG dataset. Also, hidden truncation mechanism and Bernstein density transformation strongly confirm positive sing of the risk aversion coefficient, as posterior probability $P( \alpha+E(z_j)>0|M_i,y)$ is greater than 0.98, for $i=2$ and 5.

\section{Concluding remarks}
We checked the impact of the conditional skewness assumption on the strength of the relationship between risk and expected return. On the basis of the Intertemporal CAPM model, proposed in \cite{Mert1973}, we built a GARCH-In-Mean type sampling distribution suitable in modelling such relationship. Our approach, which fully relates to the model proposed in \cite{OsiPip2000}, treats the skewness of the conditional distribution of excess returns as an alternative source of information about risk aversion. Based on the daily excess returns of the Warsaw Stock Exchange index we checked the empirical importance of the conditional skewness assumption on the relation between risk and return. Posterior inference about skewing mechanisms showed positive and decisively significant value of the coefficient of the relative risk aversion once a specific skewing mechanism was imposed in conditional Student-$t$ distribution. The greatest data support, and also very strong support of the relation postulated by Merton in \cite{Mert1973}, received skewness generated by Beta distribution transformation with two free parameters.

\bibliographystyle{plain}
\bibliography{mp_v2}

\begin{table}[ht]\scriptsize
\setlength{\tabcolsep}{0.05cm}
\centering
\caption{The conditional skewing mechanisms $p(.|\eta_i)$, defined for $u \in (0,1)$, skewness parameters $\eta_i$ and conditional symmetry restrictions in all competing specifications $M_i$, $i=1,2,3,4,5,6$.}\label{table:1}

\begin{tabular}{c | c }
\hline \hline

\begin{tabular}{c} 
$M_5$\\
Bernstein density (2 parameters)\\
$(\omega_1,\omega_2) \in (0,1)^2$, $\omega_3=1-\omega_1-\omega_2$\\
$p(u|\omega_1,\omega_2)=\sum_{j=1}^{3}\omega_jBe(u|j,4-j)$\\
symmetry: $\omega_1=\omega_2=\dfrac{1}{3}$\\
\end{tabular} 
&
\begin{tabular}{c}
 $M_1$ \\ Inverse scale factors; see \cite{FernSteel1998} and \cite{Hansen1994}\\ parameterisation as in \cite{FernSteel1998}\\
$\gamma_1>0,C=\dfrac{2}{\gamma_1+\gamma_1^{-1}}$\\ 
$I_1(u)=I_{(0,0.5)}(u)$, $I_2(u)=I_{[0.5,1)}(u)$\\
$p(u|\gamma_1)=C \dfrac{f(\gamma_1F^{-1}(u))I_1(u)+f(\gamma_1^{-1}F^{-1}(u))I_2(u)}{f(F^{-1}(u))}$\\ 
symmetry: $\gamma_1=1$ \\
\end{tabular} \\
\hline

\begin{tabular}{c} 
$M_2$\\ Hidden truncation; see \cite{Azzalini}\\
$\gamma_2 \in \mathbb{R}$\\
$p(u|\gamma_2)=2F(\gamma_2F^{-1}(u))$\\
symmetry: $\gamma_2=0$\\
\end{tabular} 
&
\begin{tabular}{c} 
$M_3$\\ Beta one parameter; see \cite{Jones2004}\\
$\gamma_3>0$\\
$p(u|\gamma_3)=Be(u|\gamma_3,\gamma_3^{-1})$\\
symmetry: $\gamma_3=1$
\end{tabular}  

\\
\hline

\begin{tabular}{c} 
$M_6$\\
The construct proposed in \cite{FerrSteel2006}\\
$\gamma_4 \in \mathbb{R}$\\
$p(u|\gamma_4)=1+l(\gamma_4)[g(u|\gamma_4)-1]$\\
symmetry: $\gamma_4=0$
\end{tabular} 
&
\begin{tabular}{c} 
$M_4$\\ Beta two parameters; see \cite{Jones2004} and \cite{JonFad2003}\\ 
$a>0, b>0$\\ 
$p(u|a,b)=Be(u|a,b)$\\ 
symmetry: $a=b=1$
\end{tabular} \\

\hline \hline
\end{tabular}

\end{table}

\begin{table}[ht]\scriptsize
\setlength{\tabcolsep}{0.1cm}
\centering

\caption{Decimal logarithms of the marginal data density values, posterior probabilities of all competing specifications $M_i$ and posterior probabilities of the positive sign of the relative risk aversion coefficient $\alpha + E(z_j)$.}\label{table:2}

\begin{tabular}{ c c c c c c c | c}
\hline \hline
 & $M_1$ & $M_2$ & $M_3$ & $M_4$ & $M_5$ & $M_6$ & $M_0$ \\
\hline \hline
log $p(y|M_i)$ & -1559.45 & -1558.50 & -1558.78 & -1558.41 & -1560.82 & -1560.10 & -1559.06 \\
\hline
$P(M_i|y), i=0, \ldots ,6 $ & 0.0353 & 0.3152 & 0.1654 & 0.3878 & 0.0015 & 0.0079 & 0.0868 \\
$P(M_i|y), i=1, \ldots ,6 $ & 0.0387 & 0.3452 & 0.1811 & 0.4246 & 0.0017 & 0.0087 & - \\
\hline
$P(\alpha+E(z_j)>0|M_i,y)$ & 0.9102 & 0.9894 & 0.9528 & 0.9972 & 0.9893 & 0.9230 & 0.9201 \\
\hline \hline
\end{tabular}

\end{table}

\end{document}